\documentclass{iau307}
\usepackage{graphicx}
\usepackage{natbib}
\usepackage{url}
\usepackage{dtklogos}
\bibpunct{(}{)}{;}{a}{}{,}

\newcommand*{\vcenteredhbox}[1]{\begingroup
\setbox0=\hbox{#1}\parbox{\wd0}{\box0}\endgroup}

\title[Spectropolarimetry: the next steps] 
{Spectropolarimetry of massive stars:\\Requirements and potential from today to 2030}

\author[G.A. Wade et al.]   
{G.A. Wade$^1$}

\affiliation{$^1$Royal Military College of Canada, Kingston, Ontario, Canada}

\pubyear{2014}
\volume{307} 
\pagerange{}
\jname{New windows on massive stars: asteroseismology, interferometry, and spectropolarimetry}
\editors{G. Meynet, C. Georgy, J.H. Groh \& Ph. Stee, eds.}

\begin{document}

\maketitle

\begin{abstract}
We develop the requirements and potential of spectropolarimetry as applied to understanding the physics of massive stars during the immediate, intermediate-term and long-term future. 
\end{abstract}

\firstsection 
\section{Introduction and scope}

Spectropolarimetry is a key technique for investigating the atmospheres, winds and envelopes of massive stars. Through the polarization of the continuum, and polarization and depolarization of spectral lines through scattering, Zeeman effect and Hanl\'e effects, polarimetry yields unique information about the geometry, density and magnetic fields of astrophysical plasma structures that are often impossible to obtain using other approaches.

In this paper we review the principal problematic themes and key questions related to massive stars that are currently being addressed using spectropolarimetric instrumentation. We then discuss forthcoming instrumentation and technical developments that will influence the field during the next 5 years, and ultimately for the period 2020-2030.

\section{Today's tools and methods}

Today, the astronomical spectropolarimetric instrumental landscape is dominated by a small number of efficient, broadband optical spectropolarimeters on 2-8m class telescopes. The principal players, summarized in Table 1, are the FORS2 multi-mode instrument on the 8m Very Large Telescope (VLT) UT1, the HARPSpol instrument at the ESO La Silla 3.6m telescope, ESPaDOnS at the 3.6m Canada-France-Hawaii Telescope (CFHT), and the Narval instrument at the 2m Bernard Lyot telescope (TBL). The basic characteristics of these instruments are summarized in Table~\ref{today}.

Today's general-purpose spectropolarimeters can be roughly classed into two categories: low resolving power instruments, such as FORS2, designed for the measurement of both continuum and line polarization with moderate precision, and high resolving power instruments, such as ESPaDOnS, designed principally to measure line polarization with high precision.

The fundamental design difference in these two categories of instruments is the situation of the spectrograph. In the low resolving power instruments, the entire polarimetric and spectroscopic unit is mounted at the Cassegrain focus of the telescope, leaving it subject to variable instrumental flexures \citep{2012A&A...538A.129B}. In contrast, the high resolution instruments employ bench-mounted, isolated spectrographs that are fed from the Cassegrain-mounted polarimetric unit using optical fibres. As a consequence, the low resolution instruments are likely limited to the detection and measurement of longitudinal magnetic fields stronger that $\sim 100$~G \citep{2012A&A...538A.129B}, whereas the high resolution instruments are capable of measuring fields of order 1 G \citep[e.g.][]{2014IAUS..302..373K,2010A&A...516L...2A}.

On the other hand, due to their fibre design the high resolution instruments struggle to measure continuum polarization accurately. As a consequence, exploitation of these instruments for studies of continuum polarization and line depolarization (relative to a polarized continuum) have been limited.

\begin{table}
\begin{tabular}{lll}
\hline\smallskip
&High resolution optical & Low resolution optical \\
\hline\smallskip
Examples & Espadons, Narval, Harpspol & FORS2, ISIS\\
Telescopes & 2-4m class  & 4-8m class  \\
Bandpass & Complete optical  & Partial optical \\
Resolving power & 65K-100K & $<2.6$K \\
Stokes parameters & IQUV, $\sigma_B\sim 1$~G & Stokes IQUV , $\sigma_B\sim 100$~G\\
Throughput & $5-20$\%  & $\sim 20$\% \\
Observational strategies & PI, LP & PI, LP\\
Lines/continuum & Lines only & Lines \& continuum\\
Analysis approach& Ind. lines, LSD & Multiline \\
\noalign{\smallskip}
\hline\hline\smallskip
\end{tabular}
\caption{Optical spectropolarimeters in the current period.\label{today}}
\end{table}

\section{Today's main themes and questions}

Much of the research performed with today's spectropolarimetric instrumentation has been focused on broadening our understanding of stellar magnetism. Magnetometry relies principally on the use of circular (Stokes $V$) spectropolarimetry, exploiting the longitudinal Zeeman effect to detect the line-of-sight component of magnetic fields in stellar photospheres. Through both PI programs and Large Programs, circular spectropolarimetry has been employed to investigate the statistics of strong, organised fields of massive stars at the late pre-main sequence, main sequence, and post-main sequence phases \citep{2008A&A...481L..99A, 2011MNRAS.412L..45P,2010MNRAS.408.2290G}. The surface geometries of magnetic fields of individual stars have been mapped in impressive detail using time series of Stokes $V$ (and sometime Stokes $Q$ and $U$) measurements interpreted using the Zeeman Doppler Imaging approach \citep[e.g.][and Fig. 1]{2011ApJ...726...24K}. In some cases, the surface field distributions have been extrapolated based on simplifying assumptions in order to model the 3 dimensional structure of the magnetic field and magnetosphere (e.g. Fig. 1).

The relationships between magnetic field characteristics and other stellar properties have been explored. A particularly energetic field of investigation has been the wind-field interaction, which has been explored from both observational and theoretical perspectives, for individual stars and populations \citep[e.g.][]{2013MNRAS.429..398P}. Extensive theoretical investigations using magneto-hydrodynamic (MHD) simulations and semi-analytic magneto-hydrostatic models \citep{2002ApJ...576..413U, 2005MNRAS.357..251T} have provided a rich theoretical context for interpretation of empirical results. 

The origin of the magnetic fields of OB stars has been a major driving question during the last decade. Analytic \citep{2010ApJ...724L..34D} and numerical \citep{2004Natur.431..819B} investigations have demonstrated that stable fossil magnetic field configurations can result naturally in stellar radiative zones from the relaxation of an initially random seed field. 

Linear polarimetry, primarily in the continuum, has been employed to investigated scattering geometries of circumstellar discs and envelopes, to probe wind structures and oblateness, and to provide independent constraints on the structure of stellar magnetospheres \citep[e.g.][]{2013ApJ...766L...9C}. Linear polarimetry of spectral lines has been exploited to a limited extent \citep[e.g.][]{2012AIPC.1429..147V,2007ApJ...667L..89H}, although useful interpretation of those results has often remained a unsolved problem.

\begin{figure}
\begin{center}
\vcenteredhbox{\includegraphics[width=12cm]{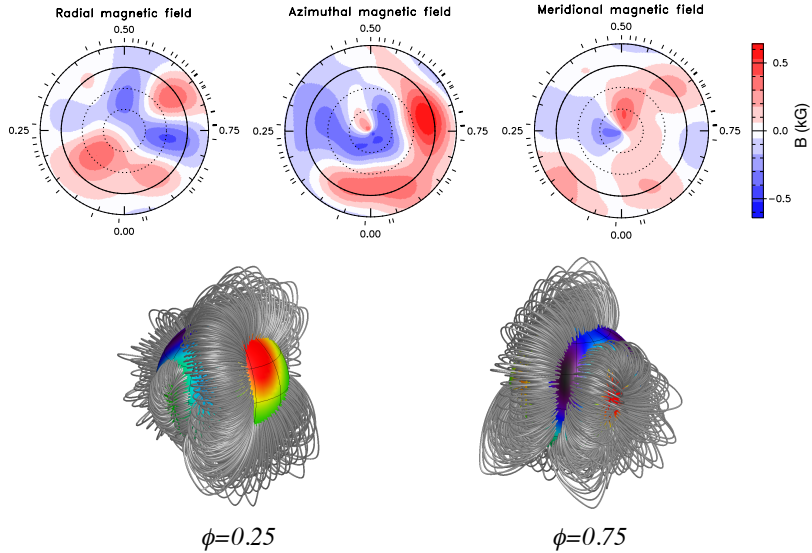}}
\vspace{-0.5cm}
\caption{{\em Top -}\ Zeeman Doppler maps of the surface magnetic field components of the magnetic B-type star $\tau$~Sco. {\em Bottom -}\ Potential extrapolation of the radial component of the surface field above the stellar surface, illustrating an idealized structure of the closed loops of the stellar magnetosphere. (Kochukhov presentation at this meeting, and priv. comm.)}
\label{fig1}
\end{center}
\end{figure}

\section{Today's limitations}

While today's spectropolarimetric technologies represent a major step forward compared to the previous generation of instruments, they are still limited in many ways that constrain our ability to address important questions. For example, while today's instruments enjoy broad bandpasses, they are all still confined to the optical window. Most instruments are designed for characterization of continuum or line polarization, but not both. The high resolution instruments are limited principally to bright stars ($V=12$ is a stretch), restricting them typically to nearby, unextincted, mainly field stars, at least at high signal-to-noise ratio. In particular, these instruments are unable to make the leap beyond the Galaxy, even to the Magellanic clouds. On the other hand, very deep spectropolarimetry of bright stars is limited by readout overheads, which are typically of the order of 1 minute per sub exposure (whereas exposure times to saturation may only be a few seconds). 

The vectorial nature of polarimetry makes is susceptible to cancellation effects. Hence most current observations are sensitive to (relatively) organised magnetic fields. Due to the thin zone of formation of the large majority of optical spectral diagnostics (i.e. in the photosphere), magnetometry is essentially confined to two dimensions (the stellar surface), as are associated models.

The large majority of investigations today rely on the interpretation of averaged peudo line profiles (e.g. Least-Squares Deconvolved, or LSD profiles). Line averaging improves precision, but the interpretation of pseudo line profiles is uncertain, leading to reduced accuracy in their interpretation.

The largest LP allocations dedicated to stellar magnetism are of order 100 nights. 

Finally, the most productive workhorse instruments (ESPaDOnS and Narval) are both located in the northern hemisphere; capabilities in the southern hemisphere are somewhat more limited.

\section{Tomorrow's questions: 2015-2020}

It seems likely that the questions to be seriously addressed during the next years will be many of those those viewed as the most challenging and exotic of today. For example, \citet{2009A&A...500L..41L} reported the detection of a very weak ($0.3$~G) longitudinal magnetic field at the surface of the bright A0V star Vega that \citet{2010A&A...523A..41P} interpreted using Zeeman-Doppler Imaging as the result of a $\sim 10$~G magnetic spot located near the star's rotational pole (Fig. 2). Additional evidence for similarly weak fields \citep[e.g.][Morel and the BOB collaboration, this conference]{2011A&A...532L..13P} has been reported in a small number of stars. Such fields may represent a much more common weak component of the magnetic field distribution of intermediate-mass and massive stars, and may help to understand the so-called "magnetic desert" and the processes leading to the observed distribution of fields in higher-mass stars. However, the detection of such fields is at the very limit of the capabilities of todays instruments. 

Investigations of the evolution of magnetic fields through stellar evolution, both before, on and after the main sequence, have been initiated during the last 10 years. However, the relative faintness of stars at the earliest pre-main sequence stages of evolution, and those located in environments very different (in terms of age, or metallicity) from the local Galactic field, has limited those studies. It is therefore likely that there will be significant pressure to extend these studies to explore characteristics of diverse stellar populations, including distant open clusters, the Galactic halo, and nearby galaxies such as the LMC and SMC (see Najarro presentation at this meeting.) 

While today's science has focused on the magnetic fields at the surfaces of stars, the next years may witness the extension of direct field constraints into stellar interiors (with the help of asteroseismology), and into winds, discs and envelopes by taking advantage of spectral diagnostics formed in these environments and present in other regions of the electromagnetic spectrum.

\begin{figure}
\begin{center}
\vspace{-1cm}
\includegraphics[width=6.9cm]{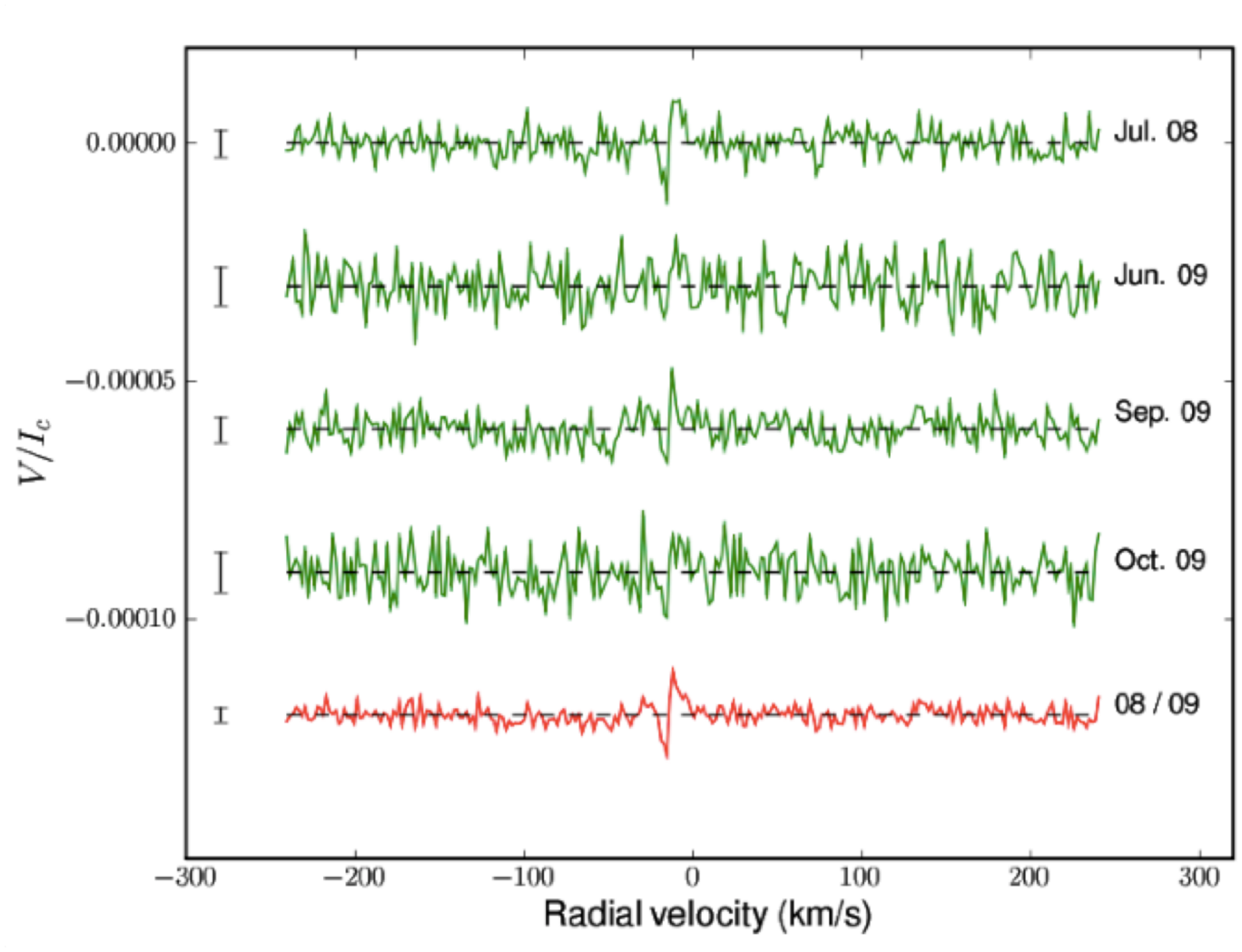}\hspace{0.2cm}\includegraphics[width=6.3cm]{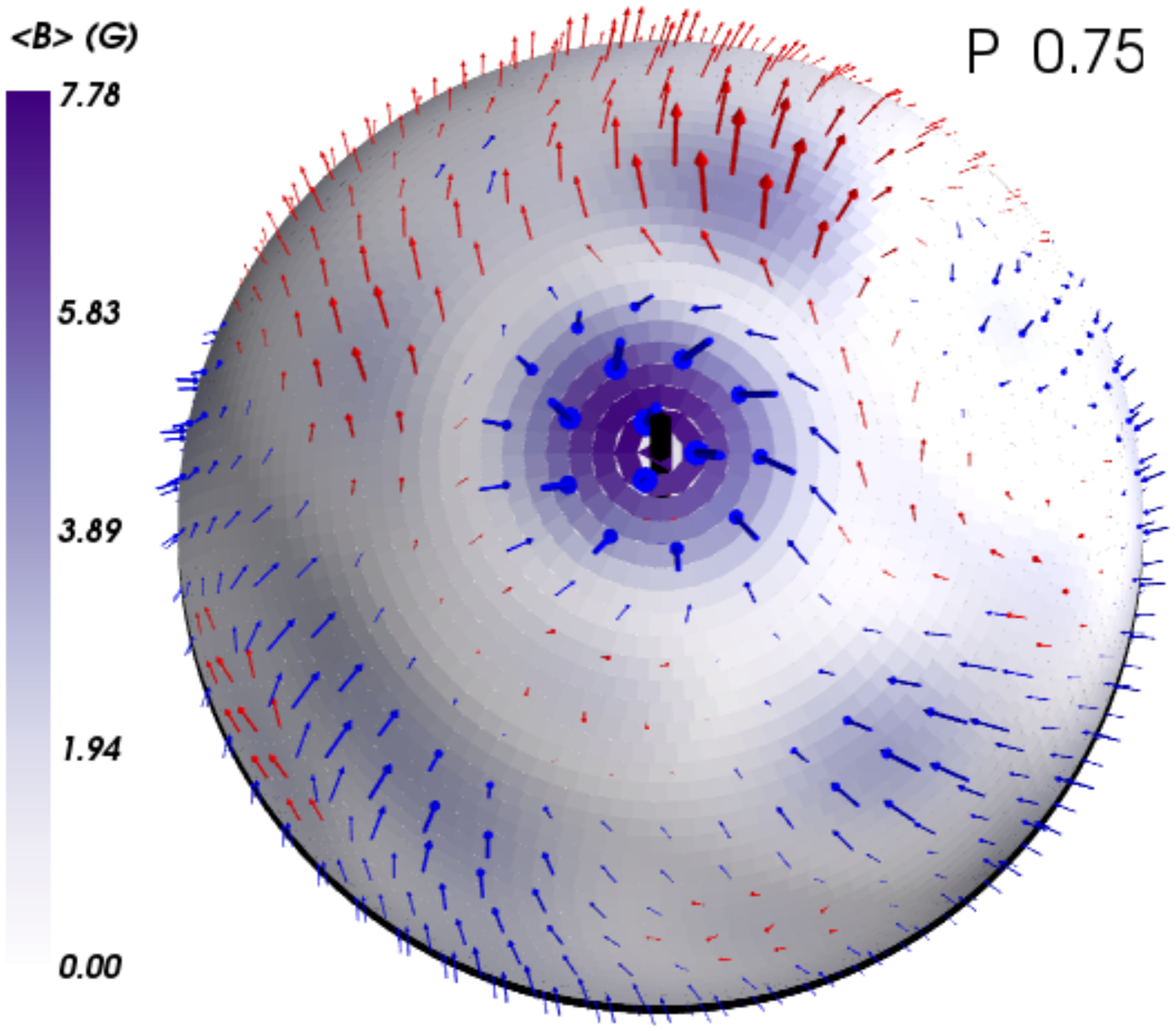}
\vspace{-1.5cm}
\caption{{\em Left -}\ Detection of a very weak Stokes $V$ signature in the mean spectral line of the rapidly-rotating A1V star Vega \citep{2009A&A...500L..41L,2010A&A...523A..41P}, corresponding to a surface magnetic field intensity of about 8~G \citep[Petit et al., submitted;][]{2014arXiv1407.3991W}. The top 4 curves represent observations obtained at different epochs with different instruments. The bottom curve is the mean. {\em Right -}\ Illustration of the complex topology of Vega's magnetic field \citep[Adapted from][]{2014arXiv1407.3991W}.}
\label{fig1}
\end{center}
\end{figure}

The demonstration that broadband linear polarization measurements with precision of the order of $10^{-4}$ or even $10^{-5}$ of the continuum flux \citep[e.g.][Faes this conference]{2013ApJ...766L...9C}, combined with increasingly sophisticated models that allow for interpretation of these data, will lead to important advances in studying of stellar rotation and oblateness (Faes presentation), wind/envelope asymmetries (Lomax and Halonen presentations), and the distribution of plasma in stellar magnetospheres (Oksala and Shultz presentations).

\section{Tomorrow's tools: 2015-2020}

A major limitation of the high resolution instruments today is their association with 4m class telescopes. Because these instruments are capable of achieving much greater magnetic precision per collected photon than the current low resolution instruments, a future priority must be to equip 8m-class telescopes in both the northern and southern hemispheres with instruments with capabilities (at least) similar to those of e.g. ESPaDOnS. For comparison, ESPaDOnS on an 8m telescope would yield a gain of 1.7 mag, corresponding to a decrease of 5x in exposure time, or 2.2x in signal-to-noise ratio. It is reasonable to imagine that existing spectroscopic instrumentation, such as UVES or X-shooter on the VLT, could be equipped with polarimeters \citep{2013ASPC..470..401S}. Alternatively, existing projects such as GRACES \citep[Gemini Remote Access to CFHT ESPaDOnS Spectrograph][]{2012SPIE.8446E..2AT} might naturally lead to the transfer of existing spectropolarimetric instruments from 4m to 8m class telescopes as new instrumentation comes online. While efforts such as these will represent major steps forward, they will still not be sufficient to permit e.g. the extension of high precision spectropolarimetry to stars outside of the Galaxy.

A major instrumental theme of the 2015-2020 era will be the introduction of high performance infrared (IR) spectropolarimeters. The two instruments that are currently in development are the "CRIRES+" upgrade to the existing CRIRES spectrograph at the ESO VLT, and the new "SPIRou" instrument under development for the Canada-France-Hawaii Telescope. The basic characteristics of these instruments are summarized in Table 2.

\begin{table}
\begin{tabular}{lll}
\hline\smallskip
&CRIRES+ & Spirou \\
\hline\smallskip
Telescopes & 8m VLT & 3.6m CFHT  \\
Bandpass & 1-2.7 $\mu$m  & 0.98-2.35 $\mu$m \\
& (single exp.) & (multiple exp.)\\
Resolving power & 100K & 75K \\
Stokes parameters & IQUV & IQUV\\
Throughput & $\sim 5$\%  & $\sim 15$\% \\
Commissioning & 2017 & 2017\\
\noalign{\smallskip}
\hline\hline\smallskip
\end{tabular}
\caption{IR spectropolarimeters in the period 2015-2020.}
\end{table}

\subsection{CRIRES+}

CRIRES+ represents a major upgrade to the current CRIRES (Cryogenic high-resolution InfraRed Echelle Spectrograph) installed at the VLT \citep{2014arXiv1407.3057O}. This transformation will see many improvements to the existing AO-assisted instrument, while retaining its industry-leading ability to capture long-slit, high-resolution ($R\sim 100$K) spectra over a wide spectral range (0.95-5.2 $\mu$m).  The most significant upgrade is the introduction of a cross-disperser unit and larger detectors that will increase the single-exposure wavelength coverage by about a factor of 10 compared to the existing setup. Furthermore, CRIRES+ will see the introduction of polarimetric optics that will allow for the high-precision measurement of circular (Stokes $V$) and linear (Stokes $Q$ and $U$) polarization within spectral lines in the range of 1-2.7 $\mu$m. In addition to these significant changes, CRIRES+ will undergo many other smaller upgrades that will enable it do new and exciting science: the search for super-Earths in the habitable zone of low-mass stars; the atmospheric characterization of transiting planets; the origin and evolution of stellar magnetic fields. CRIRES+ will remain a general-purpose user facility and will offer comprehensive calibration and data-reduction support. Commissioning of CRIRES+ is projected for 2017.

\subsection{SPIRou}

SPIRou is a near-infrared spectropolarimeter / velocimeter currently under construction for CFHT. SPIRou aims in particular at becoming the world-leader on two forefront science topics, (i) the quest for habitable Earth-like planets around very- low-mass stars, and (ii) the study of low-mass star and planet formation in the presence of magnetic fields. In addition to these two main goals, SPIRou will be able to tackle many key programs, from weather patterns on brown dwarf to solar-system planet atmospheres, to dynamo processes in fully-convective bodies and planet habitability. The science programs that SPIRou proposes to tackle are forefront (identified as first priorities by most research agencies worldwide), ambitious (competitive and complementary with science programs carried out on much larger facilities, such as ALMA and JWST) and timely (ideally phased with complementary space missions like TESS and CHEOPS). 
SPIRou is designed to carry out its science mission with maximum efficiency and optimum precision. More specifically, SPIRou will be able to cover a very wide single-shot nIR spectral domain (0.98-2.35 $\mu$m) at a resolving power of about 75K, providing unpolarized and polarized spectra of low-mass stars with a $\sim 15$\% average throughput and a radial velocity (RV) precision of 1 m/s. Commissioning of SPIRou is projected for 2017.

\begin{figure}
\begin{center}
\vcenteredhbox{\includegraphics[width=5.5cm]{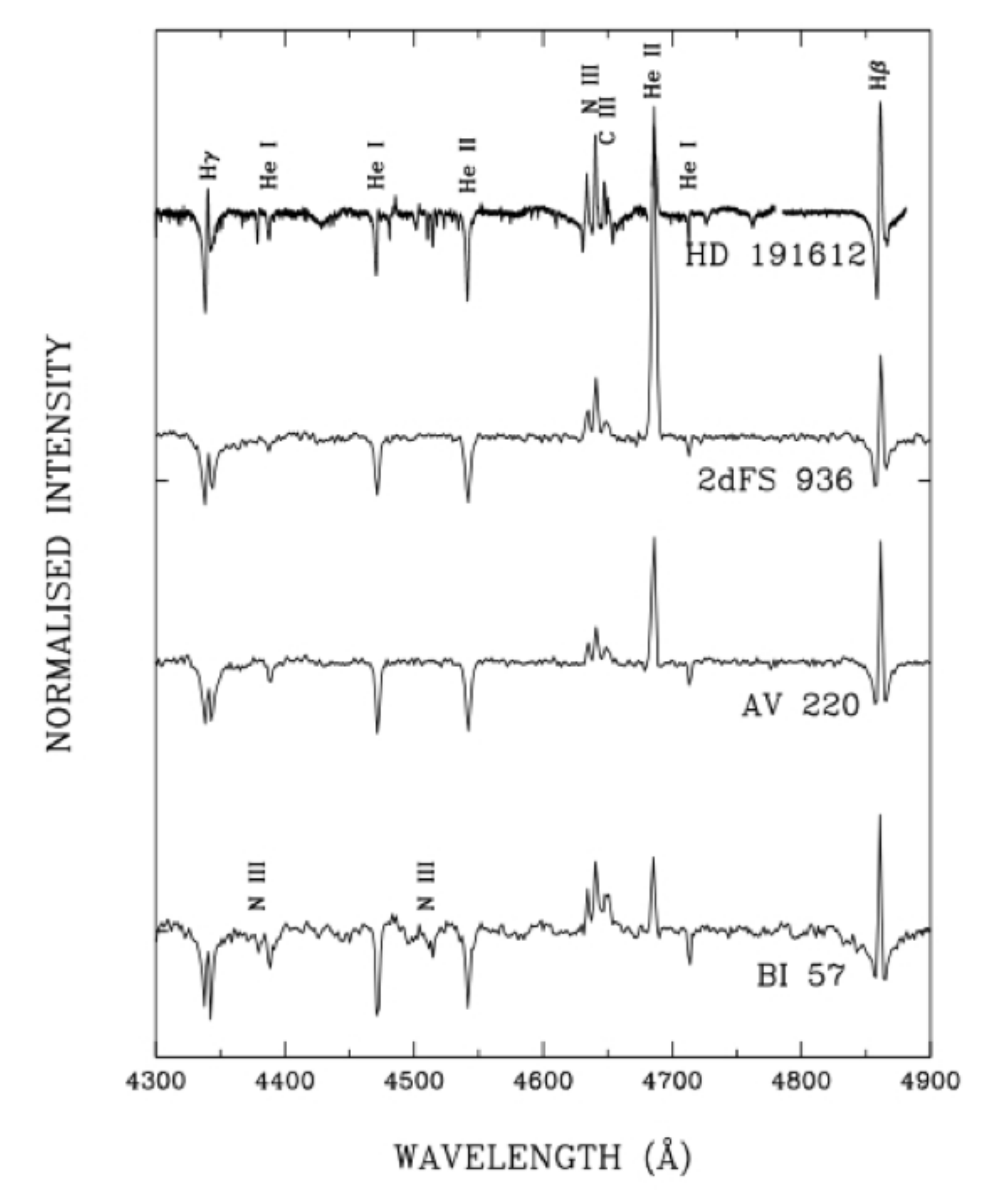}}\vcenteredhbox{\includegraphics[width=8cm]{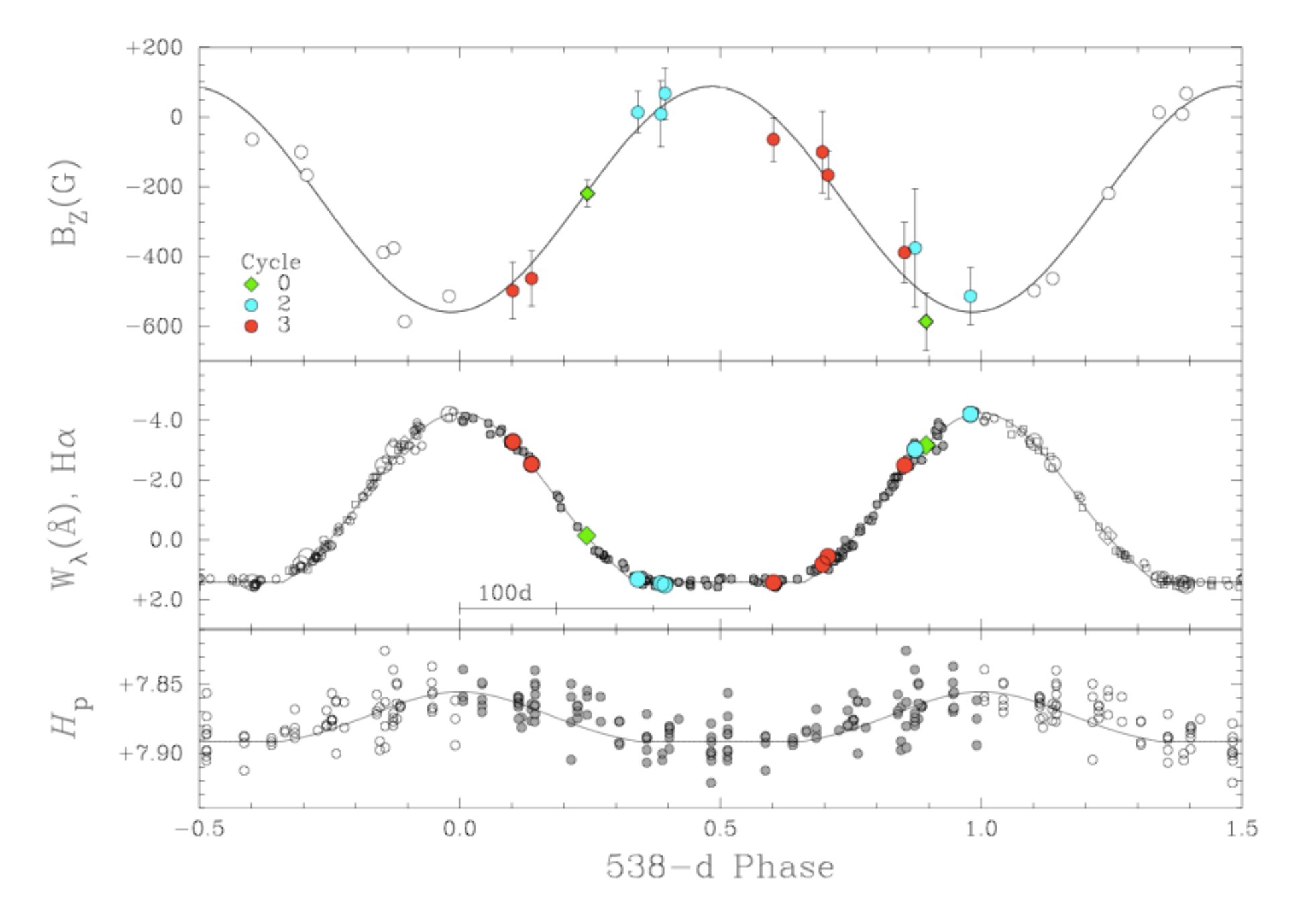}}
\caption{{\em Left -}\ Optical spectra of Of?p stars identified in the LMC and SMC \citep[][Howarth et al., in prep]{2000PASP..112...50W,2001ApJ...550..713M,2004MNRAS.353..601E}, as coma red to the Galactic Of?p star HD 191612. {\em Right -}\ Illustration of the variations of the magnetic field, H$\alpha$ equivalent width and Hipparcos magnitude of the Galactic Of?p star HD 191612 \citep{2007MNRAS.381..433H,2011MNRAS.416.3160W}.}
\label{fig2}
\end{center}
\end{figure}


\section{Around the bend: 2020-2030}

\subsection{Extremely large telescopes}

The defining tools of ground-based astronomy in the 2020-2030 timeframe will be the so-called "giant" telescopes - the 39m European Extremely Large Telescope (E-ELT), the 30m Thirty Metre Telescope (TMT), and the 24.5m Giant Magellan Telescope (GMT). The former two facilities will be located in the southern hemisphere (northern Chile), while the latter will be in the North (Hawaii).

Equipping any of these telescopes with a moderate resolution ($R\gtrsim 10000$) fibre fed optical spectropolarimeter would be an utter game-changer, yielding an improvement of 5 magnitudes relative to existing high resolution instrumentation, with minimal reduction in polarimetric or magnetic precision. This translates into a gain of 10x in signal-to-noise ratio, or 100x in exposure time. Such an instrument could potentially achieve a magnetic detection precision of a few hundredths of a G, routinely measure Zeeman Stokes $QUV$ polarization in individual spectral lines of typical magnetic O stars similar to HD 191612 (see Fig. 3) as faint as 10th magnitude, and detect fields comparable to that of HD 191612 in early O stars in the Magellanic clouds. As candidate extra-Galactic magnetic O stars have already been identified spectroscopically (see Fig. 3), this capability is a natural next step in understanding the environmental factors influencing massive-star magnetism.

Unfortunately, there is currently no high resolution spectropolarimeter planned for any of the giant optical telescopes.

\subsection{UVMag}

UVMag \citep[][and Coralie Neiner's presentation at this meeting]{2014Ap&SS.tmp..292N} is a planned intermediate-sized space telescope (1.3 m) supporting optical and UV spectropolarimetry with spectral resolution sufficient for Zeeman Doppler surface imaging. UVMag would enable continuous, high cadence line and continuum spectropolarimetry from 117-900 nm at high ($\sim 25,000$) resolving power. The project is currently supported by France for an ESA M-class mission (under the mission name "Arago") with planned launch in the late 2020s. The unique observations that UVmag will provide will allow a simultaneous and continuous view of stellar environments from the deep photosphere to the peripheries of their winds, discs and magnetospheres, potentially exploiting signal from scattering, longitudinal and transverse Zeeman effects, and Hanl\'e effect.

Recent spectral synthesis calculations by Colin Folsom (Fig. 4) demonstrate the power and richness of the UV domain for detection and characterization of fields in hot, massive stars.

\begin{figure}
\begin{center}
\vcenteredhbox{\includegraphics[width=6.7cm]{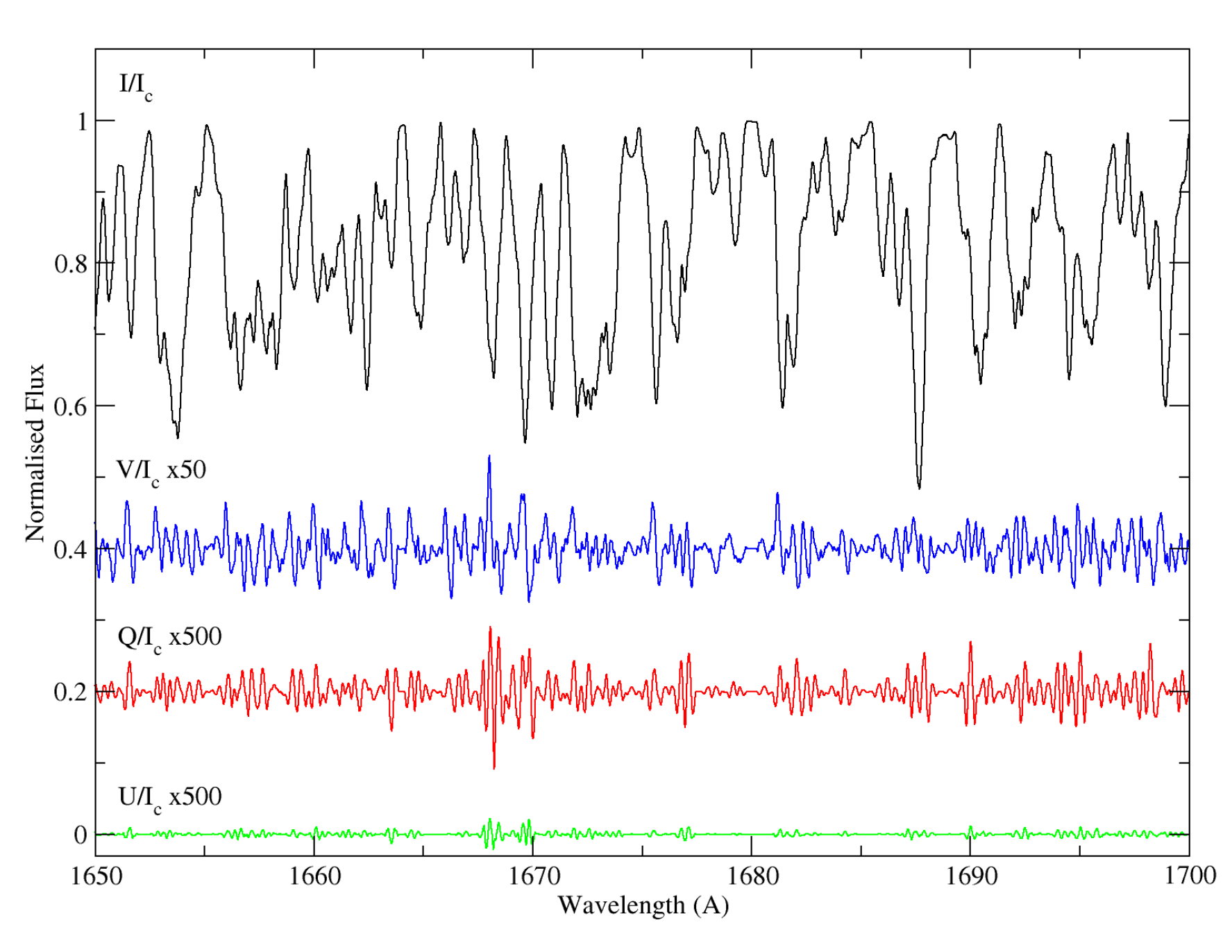}}\vcenteredhbox{\includegraphics[width=6.7cm]{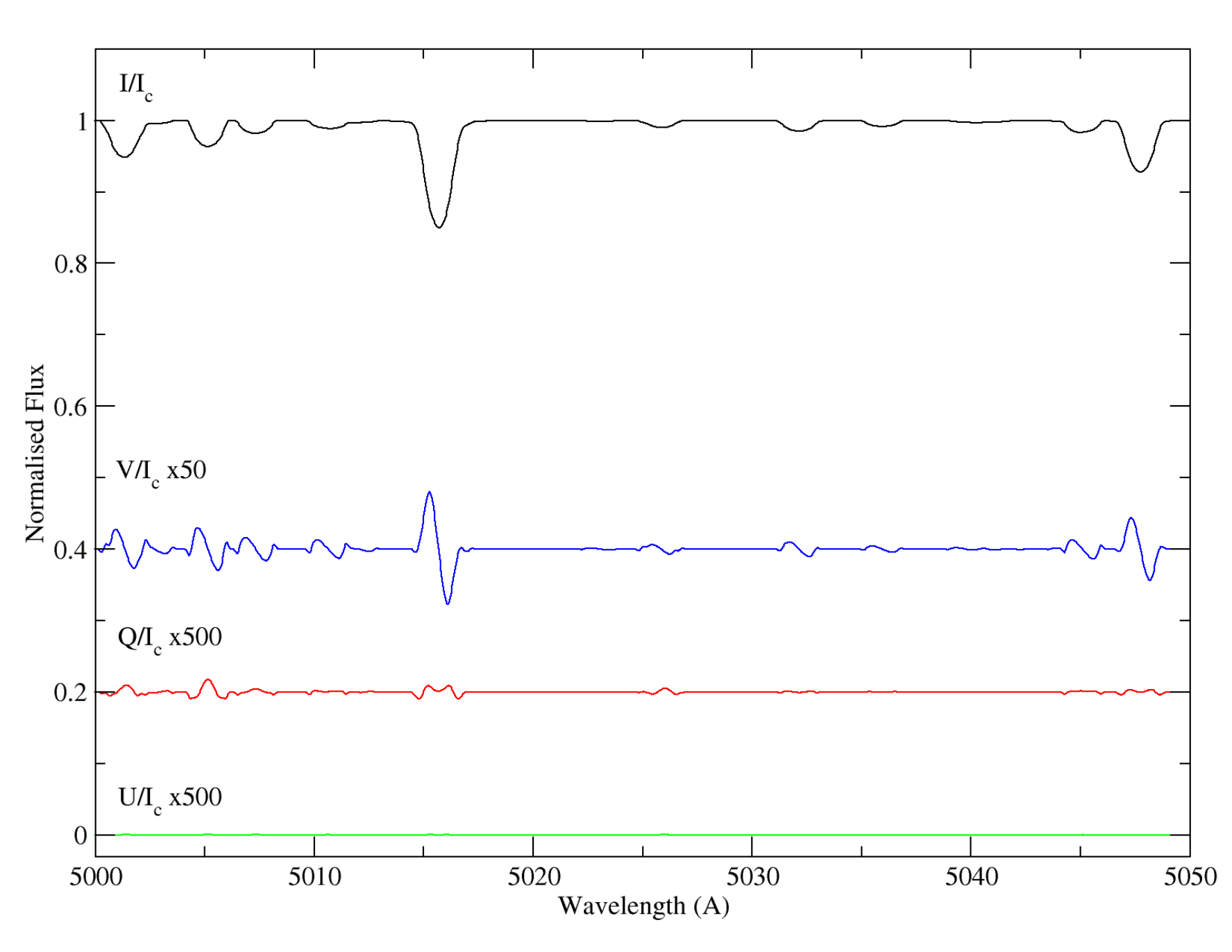}}
\caption{Spectral regions of width 50~\AA\ computed in all 4 Stokes $IQUV$ parameters for a B1V star with a 1~kG dipolar magnetic field in its photosphere (Folsom et al., priv. comm.). {\em Left -}\ Ultraviolet. {\em Right -}\ Optical. Note both the richness and the strength of the UV polarimetric signal. }
\label{fig1}
\end{center}
\end{figure}

Simultaneous with these new ground and space technologies, new techniques, strategies and modeling approaches must be developed (e.g. Spectro Polarimetric Interferometry \citep[SPIn][]{2003SPIE.4843..484C}, large-scale, holistic models combining asteroseismology and spectropolarimetry, leading to sophisticated, realistic predictions). Perhaps new observational strategies will be developed, with different scales of large programs. 

\section{Conclusion}

In conclusion, today's suite of front-line spectropolarimetric instrumentation retains significant potential for scientific productivity and discovery. During the next 5 years, we will witness the introduction of powerful infrared spectropolarimeters as the leading instruments. At the same time, commissioning of high resolution optical instruments on 8m class telescopes should be a priority. In the era 2020-2030 we may encounter the first space spectropolarimeter working at both optical and ultraviolet wavelengths. We will witness ground-based optical astronomy dominated by a new era of giant telescopes. However, there is currently no plan to equip any of these observatories with high resolution spectropolarimetric equipment.


\bibliographystyle{iau307}
\bibliography{IAUS307_Wade_prospective}

\begin{discussion}

\discuss{Khalack}{Could you, please, comment on the use of the LSD method for UV space spectropolarimetry in the region 117-400 nm where a lot of lines is strongly blended?}

\discuss{Wade}{Colin Folsom's calculations using synthetic spectra and realistic atomic data suggest the outlook is favourable!}

\end{discussion}

\end{document}